# Challenges for Security Assessment of Enterprises in the IoT Era

**Yael Mathov**
Department of Software and
Information Systems Engineering
Ben-Gurion University, Israel

**Noga Agmon**
Department of Software and
Information Systems Engineering
Ben-Gurion University, Israel

**Asaf Shabtai**
Department of Software and
Information Systems Engineering
Ben-Gurion University, Israel

**Rami Puzis**
Department of Software and
Information Systems Engineering
Ben-Gurion University, Israel

**Nils Ole Tippenhauer**
CISPA Helmholtz Center
for Information Security
Germany

**Yuval Elovici**
Department of Software and
Information Systems Engineering
Ben-Gurion University, Israel

## Abstract

For years, attack graphs have been an important tool for security assessment of enterprise networks, but IoT devices, a new player in the IT world, might threat the reliability of this tool. In this paper, we review the challenges that must be addressed when using attack graphs to model and analyze enterprise networks that include IoT devices. In addition, we propose novel ideas and countermeasures aimed at addressing these challenges.

## 1 Introduction

The adoption of Internet of Things (IoT) technology and integration of IoT devices into the networks of enterprise organizations has increased dramatically worldwide [1]. Unfortunately, many IoT devices manufacturers focus on novel functionalities and short time-to-market, while ignoring the security and privacy risks [2]. As a result, these IoT devices are deployed with a variety of unresolved vulnerabilities that may be exploited by attackers and thus introduce new risks to the organizations.

Traditional security and IT management tools used in enterprise organizations were primarily designed for networks with static hosts (e.g., servers and PCs). The special features and characteristics of IoT devices have been overlooked, and therefore, existing security tools and methods do not address them. Although scanning for vulnerable IoT devices can be performed with various scanning tools [3], they might be less effective for IoT devices in larger enterprises networks (e.g., in this setting they might be unable to identify the devices or vulnerabilities within a reasonable amount of time or supply partial output).

IoT devices introduce new challenges in terms of cyber security. First, the diversity of the communication medium and protocols used by IoT devices, especially short-range communication like Bluetooth or ZigBee, might open an unmonitored attack vector for IoT devices. As has been demonstrated, a compromised smart lamp could infect similar devices via physical proximity (via ZigBee protocol) and render network isolation useless [4]. In addition, the mobility of some IoT devices and the dynamic changes in network topology caused by the devices, impair the development of efficient security solutions [5]. Moreover, IoT devices affect each other in different ways, both explicitly (an application that uses multiple devices) and implicitly (a smart light bulb could trigger a smart light sensor). Therefore, there is a need to address the new security challenges posed by IoT devices by adapting existing security assessment methodologies and remedies.

In this paper, we review the security assessment challenges in enterprise networks containing IoT devices. Specifically, we focus on attack graphs as an important tool for security assessment that needs to be adapted to the new reality of IoT technology. We discuss the gaps and limitations of the existing security assessment methodologies relying on attack graphs. To the best of our knowledge, existing research on attack graphs does not consider the unique behavior of IoT devices and may fail to deliver reliable results in enterprise environments that contain IoT devices.

This paper lists issues that must be considered when employing attack graphs to model and analyze networks that include IoT devices, and presents possible ways of resolving these issues. Specifically, we discuss an IoT scanner tailored to collect data specific to IoT devices [6]. We further propose a novel idea for adjusting attack graphs to address the challenges related to the mobility of IoT devices and dynamic network changes. Finally, we demonstrate an application of the adjusted attack graphs to determine the deployment of a set of IoT devices which minimizes the security risk to the organization, and map directions for future research.

We believe that this work will help organizations and researchers better understand the unique challenges and key limitations of performing attack graph analysis in the presence of IoT devices. Furthermore, we provide a concise list of changes that must be made in order to enable attack graph analysis on such networks.

## 2  Background on Attack Graphs

An attack graph is a model of a computer network that encompasses computer connectivity, vulnerabilities, assets, and exploits. Attack graphs are used to represent collections of complex multi-step attack scenarios which traverse an organization from an initial entry point to the most critical assets. By analyzing the attack graph, a security analyst can assess the risks of potential intrusions and devise effective protective strategies. The attack graph analysis methodology contains three main stages:

- Network and vulnerability scanning.
- Attack graph modeling.
- Attack graph analysis.

### 2.1  Network and Vulnerability Scanning

Most attack graphs rely on computer network configuration, such as the network topology, firewall configuration, operating systems, open services, etc. [7] Therefore, different tools and databases are used to improve the scanning and mapping process. For instance, Nessus [8] and OpenVAS [9] are vulnerability scanners commonly used to map the systems, services, and vulnerabilities within an organization. Vulnerability databases, such as the National Vulnerability Database (NVD) [10] and the Open Source Vulnerability Database (OSVDB) [11], also contain additional information about the vulnerabilities (attack complexity, required privileges, exploit severity, etc.) used to augment an attack graph.

Vulnerability scanners are also used to assess the connectivity between different network segments within an organization. However, these tools do not capture connectivity between individual computers allowed through exceptions defined in the firewalls. Therefore, additional processing of firewall configurations is necessary to complement the network connectivity information required for accurate attack graphs.

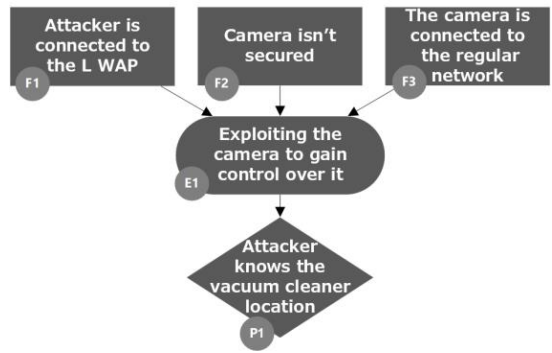

Figure 1: Simple attack graph.

## 2.2 Attack Graph Modeling

State-of-the-art attack graph modeling tools, such as MulVAL [12] [13] and NetSPA [14], use logical expressions to represent the possible attacker actions with their preconditions and acquired assets. Other graph-based modeling techniques, such as ADversary VIew Security Evaluation (ADVISE) [15] and Attack-Defense Trees (ADTrees) [16], aim to provide a useful security assessment solution. The IoT challenges discussed in this paper are relevant to all attack graph modeling techniques, but the examples provided are based on the logical attack graphs.

MulVAL relies on the Datalog programming language for describing the attack graph. It defines three types of nodes:

- Fact nodes, which describe the initial network condition (connectivity and vulnerabilities), are usually drawn as rectangles in typical visual representations of attack graphs (see Figure 2).
- Exploit/action nodes (visually denoted as ovals) represent the acts of exploiting a vulnerability or acquiring an asset.
- Privilege nodes (visually denoted as rhombuses) represent assets (e.g., information, access privileges) obtained by the attacker as a result of performing an action.

In order to obtain a privilege (P1), an attacker must execute one of the actions (E1) leading to the privilege node (logical OR). To use an exploit (E1) the attacker needs all the privileges and facts (F1, F2, F3) that lead to the exploit (logical AND). Once executed, the exploit node implies all the post-conditions (privileges) it leads to.

## 2.3 Attack Graph Analysis

Attack graphs are well formed planning problems, due to their logical AND-OR structure [17]. State-of-the-art planners can be used to determine the optimal attack plans, which in turn can be used to assess the probability of a successful attack, the expected cost of an attack, the assets at highest risk, etc. These are important capabilities, since patching all the vulnerabilities in a network is not always possible (e.g., due to budget limitations or operational constraints). Therefore, although scanners can help detect vulnerabilities in the system, they do not serve as a standalone solution for securing a network. Attack graphs can evaluate the system based on different configurations and determine the optimal patching or deployment strategy. Furthermore, they can also be used to improve intrusion detection systems (IDSs) by correlating alerts that correspond to exploits along feasible attack plans.

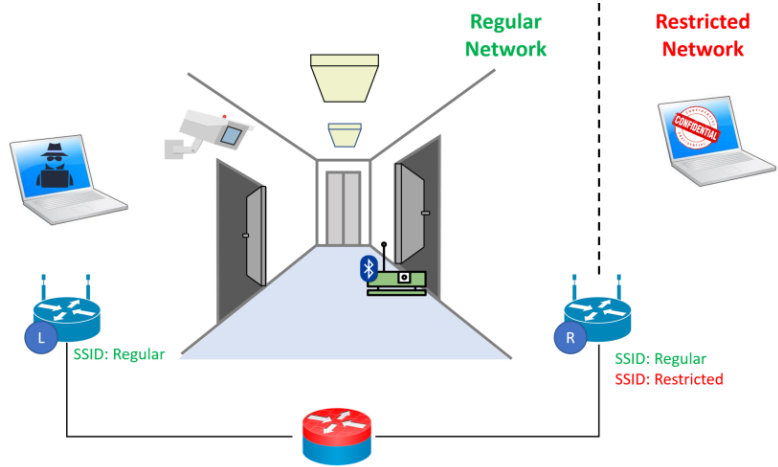

Figure 2: An example of an enterprise network with two IoT devices: IP camera and robotic vacuum cleaner. Two VLANs are defined: Regular (green) and Restricted (red). An attacker controls a host in the room on the left which is connected to VLAN Regular via L.

## 2.4   Known Limitations of Attack Graphs

Two major limitations of attack graphs acknowledged in the literature are scalability and dynamic adjustments [18]. Attack graphs do not scale well. Networks with dozens of computers result in attack graphs with thousands of nodes. Furthermore, probabilistic attack graphs might require solving huge Markov decision process (MDP) problems which is infeasible even for small computer networks.   Proposed solutions attempt to reduce the size of the graph or split it into multiple independent problems. According to [18], an adequate attack graph representation does not yet exist. Dynamic adjustment of elements in the network limits the use of attack graphs in dynamically changing networks. This problem is relatively new in the literature, and as with the above-mentioned scalability issue, an effective solution remains elusive.

## 3   New Challenges Introduced by IoT Technology

In this section, we discuss the new challenges, in each stage of the attack graph methodology, which must be considered and resolved when using attack graphs for modeling and analyzing networks with IoT devices. We will use the following scenario with two IoT devices (see Figure 2) to demonstrate the challenges and potential solutions throughout the paper.

The scenario includes an enterprise network with two IoT devices: security camera and smart vacuum cleaner, and two VLANs: Regular with permissive security policies (green) and Restricted with hosts containing confidential data (red). Two wireless access points (WAP) are located on two sides of a corridor: left (L) and right (R). Access to the Regular network is provided on both sides of the corridor. Access to the Restricted network is provided only in the room on the right side of the corridor and requires password authentication. A host controlled by the attacker is located in the room on the left and is connected to VLAN Regular via confidential data, and although she knows the credentials for the Restricted network, she has no physical access to it. A smart vacuum cleaner cleans the room on the left side of the corridor in the morning and the room on the right side of the corridor at night while connecting to the VLAN Regular, via L or R respectively. The vacuum cleaner has an unsecured Bluetooth capability that allows the attacker to infect the device when it is located in the left side of the corridor (due to physical proximity). A security camera is connected to VLAN Regular. The location of the camera is variously on the left or right side of the corridor and uses the closest WAP to access the network. Regardless of the position of the security camera, it monitors the whole corridor including the movements of the vacuum cleaner.

### 3.1 Scanning for Connectivity and Vulnerabilities

The first stage of security assessment of an organization includes mapping the network components, hosts, and services. Therefore, the goal of the scanning process is to collect all the essential data in order to derive a reliable and complete model of the organization (i.e., the nodes and edges of the attack graph).

Standard scanning tools rely on fingerprints or heuristics tailored to specific operating systems, services, applications and vulnerabilities. The *diversity* of new IoT devices, their services and configurations poses a serious challenge to development of efficient scanning tools capable of identifying IoT devices within the organizational network. Moreover, many IoT devices have low resources, and as a result, installation of *monitoring* software is not always possible for smart devices. Thus, scanning the network with commonly used tools results in only partial information, which in turn, leads to fragmentary attack graphs and inaccurate security risk assessment.

Short-range communication protocols could be exploited as part of a cyber-attack aimed at subverting IoT devices. Scanning of the various short-range communication channels used by IoT devices requires technology that is not readily available in commercial off the shelf scanners. Moreover, certain low power IoT devices occasionally disable their communication to save battery power. Therefore, it would be challenging to identify a *cross-device communication capability* of a new device if it keeps silent during the scan. Partial mitigation of this challenge may include mapping the *physical location* of each IoT device along with the physical locations of regular IT devices (PCs and servers). The physical proximity between devices may suggest the presence of a short-rang connection between the devices. This method is inaccurate, and scanning short-range communication needs to be addressed in a new way.

In contrast to IoT devices which are frequently upgraded or replaced, regular IT equipment like PCs or servers is rarely replaced by newer equipment and is also more stable. Smart IoT devices are *rapidly changed* rendering the results of past scans irrelevant after a short time period. However, the network changes that are caused by IoT devices usually result from the devices' unique design. Unlike PCs and servers, *mobile* IoT devices change their location and their connection points within the organizational network. For example, the vacuum cleaner in Figure 2 may connect to the Regular VLAN via two different WAPs. In addition, other *devices capabilities* can change the connectivity; an IoT device may behave differently according to changes in its environment that are monitored by the device's sensors. For instance, a security camera may be configured to start recording and enable the streaming service only when movement is detected by its motion sensor; otherwise the streaming service port will be closed for security reasons. In order to create a comprehensive model of the network, the scanning process should rapidly identify changes in the IoT devices' connectivity and enabled services.

### 3.2 Modeling

The next stage of creating and analyzing attack graphs is attack graph modeling.

Most attack graph models allow defining connectivity between specific defines, and therefore, short-range communication do not pose a serious challenge in this stage. However, it still need to be addressed. The *mobility* capability of some IoT devices primarily affect the modeling of short-range wireless communication (which is based on physical proximity). Therefore, the *physical location* of hosts need to be represented in the attack graph to improve the security in the face of the changes created by the new technology.

Some IoT devices' behavior is based on information that is collected via sensors and the *feedback from other IoT devices* in the network. As a result, two identical devices may act completely different when placed in different areas. Moreover, those *unique capabilities* could be exploited by an attacker. For instance, the attacker can gain information by tapping a camera's video stream and remotely monitoring an organization. For that reason, in addition to the network connectivity, the special functionalities of each IoT device in the network is important for security assessment.

Table 1: IoT security challenges and their effects on different stages of creating an attack graph of a network that contains IoT devices.

| Challenge | Description | IoT Attack Graph | Stage |
|---|---|---|---|
| **Diversity** | It is challenging to find a general solution for the wide variety of IoT devices. Moreover, devices of the same type might behave differently. | Identifying an IoT device, its connectivity, list of vulnerabilities, or even its type is a challenging task. | Scanning |
| **Lack of monitoring** | IoT devices and some of the protocols they use are monitored only partially or not at all. | Collecting information on IoT devices is more difficult due to the lack of monitoring of the unique device protocols. | Scanning |
| **Cross device capabilities** | IoT devices may implicitly and indirectly affect other devices and applications. The dependencies between the devices are hard to correctly assess. | Cross device dependencies must be known and represented in the attack graph. Therefore, the capabilities of each device need to be represented in the graph according to the current context of the devices in the network. | Scanning, Modeling, Analysis |
| **Physical location** | Cyber-attacks might spread via physical proximity as well as via an Internet connection. | Physical locations of hosts and devices need to be scanned, measured and modeled in the attack graph in order to create a reliable representation of short-range communication protocols. | Scanning, Modeling |
| **Rapid change** | The behavior of IoT devices may change rapidly. Furthermore, IoT devices are added and replaced quicker than traditional hosts. | Snapshots of a network with IoT devices become irrelevant quicker than snapshots of a network without such devices. | Scanning, Modeling, Analysis |
| **Mobility** | Some IoT devices have the capability to change their location or connection state. | Scanning the device's location and connectivity over time might lead to different results. Furthermore, temporal aspects need to be modeled in the attack graph. | Scanning, Modeling, Analysis |
| **Device capabilities** | IoT devices usually have unique capabilities that could be manipulated during a cyber-attack. Traditional monitoring and security tools might be limited when used for IoT devices with novel functionality and sensors. | Devices' capabilities and the effect of each device on the network need to be scanned and modeled in the attack graph. | Scanning, Modeling, Analysis |
| **Quantity** | The number of connected IoT devices is increasing each year. | The number of connected IoT devices affects the size and complexity of the scanned network and respective attack graph. | Modeling, Analysis |

IoT devices may *change their connectivity frequently* over time due to unique capabilities, and add temporal aspect to the network. For example, the vacuum cleaner in Figure 2 cleans distinct areas at different times of the day. Therefore, the attack graph must represent the network in the morning and at night, as well as the transition period between the two times of day. If the temporal aspect is not part of the attack graph, some threats might be missing from the graph.

Scaling still needs to be addressed in order for large enterprises to use the attack graph assessment tool, and this is exacerbated when IoT technology is added to the picture. In the near future, the *increasing number of connected devices* could cause an increase in the scale of every network. Therefore, it is even more important than ever to address the scaling challenge for attack graphs if this tool is going to continue to be relevant.

### 3.3 Analysis

Finally, the attack graph of the organization's network is analyzed using different techniques in order to identify threats, security holes, etc.

In addition to modeling temporal aspect in attack graph, there are also implications on analyzing those aspects in the graph when IoT devices are involved. The state of each node might change between the temporal intervals and need to be addressed in the analysis stage. For instance, if the vacuum cleaner in Figure 2 gets infected in the morning by an attacker's malware when cleaning the room on the left, the attack graph for the next night might look different from the one from the night before. The analysis process need to support the location changes over time, that caused by the device's mobility. Therefore, the temporal aspect should be considered when determining the state of each graph node as well.

The *rapid changes* in the network due to the behavior, *capabilities*, and *cross device capabilities* of IoT devices can be challenging for an assessment tool such as an attack graph, which has traditionally designed for static networks. Analyzing a snapshot of a network that contains IoT devices may not provide a real-time solution which monitors and represents the network's modifications. However, developing a real-time attack graph is not *scalable* and might create additional challenges to those presented in this paper.

Table 1 summarizes the security challenges related to IoT devices discussed in this section, along with their impact on the traditional attack graphs used for security assessment.

## 4 Adjustments of Attack Graphs

In this section, we present novel ideas for future research aimed at addressing and overcoming some of the previously mentioned challenges.

### 4.1 Monitoring of Wireless Traffic

Due to the distributed and wireless character of IoT communications, traffic monitoring could also leverage passive observations on the wireless link layer. For example, systems such as IoTScanner [6] could be used to obtain real-time information about current network topologies and estimate the types of the end devices involved [19].

Such monitoring tools utilizes several radios for wireless standards to passively collect traffic in the neighborhood. All the traffic collected is then processed based on a set of rules to extract features for both real-time analysis (e.g., device detection, current volume of traffic) and analysis performed at a later time (e.g., for device classification). Network connectivity graphs for the different wireless standards are continuously produced as a result of the real-time analysis. The information could then be stored in suitable graph-based data structures and used to enrich the attack graph-based assessment as described in this work. Alternatively, the monitor could be used to detect configuration and connectivity changes, and trigger appropriate actions.

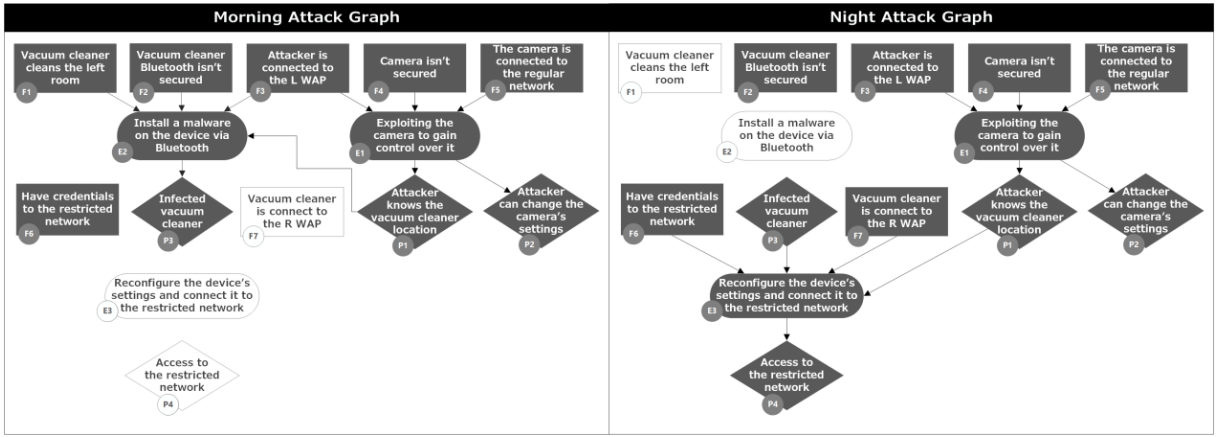

Figure 3: Example of a temporal attack graph. The colors represent the state of each node: a white node is "off" while a gray node is "on."

### 4.2 Temporal Attack Graphs and Node State

A temporal attack graph that represents the model of a network at different times of the day could provide a possible solution for the dynamism of a network. The temporal attack graph that we present is a dynamic attack graph which represents a network in each predetermined temporal. For a temporal interval, the graph is based on nodes in one of two possible states: the ``on" state when the node needs to appear in the graph for this interval, or the ``off" state if it does not need to appear in the graph.

In Figure 3, we demonstrate how a temporal attack graph can model two temporal intervals (morning and night) of the network presented in Figure 2. In the morning (left), when the vacuum cleaner is on the left side of the corridor, the attacker may infect the vacuum clearer by exploiting the physical proximity and the vulnerable Bluetooth channel. However, the attacker can not use the device to connect to the Restricted VLAN since she does not have physical access to R. In the morning (left) this constraint is depicted by the unsatisfied precondition F7 of action E3 (since F7's state is "off"). As a result, there is no path that leads the attacker to her goal. However, at night (right), the vacuum cleaner is located on the right side of the corridor and is connected to the Regular VLAN via R (precondition F7's state is "on"). If at the same time the vacuum cleaner is infected, then the attacker is able to execute action E3 and eventually reach her goal.

In addition to the constant state of the nodes in a traditional attack graph, (where the node always appears), the state of a node might be updated according to changes in the networks. A volatile behavior in the network is represented in the temporal attack graph as a change in the state of a node over time. For example, the vacuum cleaner's location in the morning turns "on" F1 and turns "off" F7. At night, the states change again but in the opposite direction. A persistent behavior in the network may change a node's state from "off" to "on" (and vice versa) starting from a certain event (e.g., P1 in the morning will leave the compatible privilege "on"). Furthermore, the state of each node may affect the state of other nodes. For instance, if one of the preconditions for an action node is "off," so is the action.

Such temporal attack graph could be implemented in several ways. Modeling the temporal attack graph is based on the results of the scanning process over time and therefore, represents the changes in the network. However, adding the time aspect to the attack graph might increase its complexity, and as a result, the development of a temporal attack graph must overcome the scaling challenge.

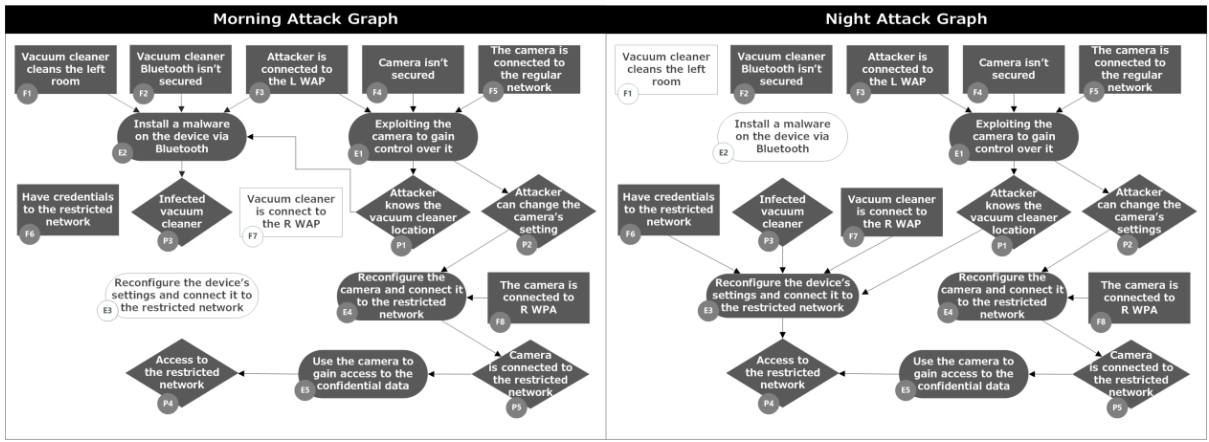

Figure 4: Example of an attack graph for the security camera. In the upper part of the figure, the camera is deployed on the right side of the corridor. There is attack path from the attacker to the confidential data (the goal). In the lower part of the figure, the camera is deployed on the left side of the corridor. There is no path for the attacker to get access to the confidential data.

### 4.3 Capabilities

Each IoT device contains the sensors and functionality it needs to fulfill its task. However, this legitimate capability could be exploited for malicious porpoises.

The representation of IoT devices capabilities as nodes is demonstrated in Figure 3 in node P1. An attacker can exploit the security camera to gain access to the video stream and learn the location of the smart vacuum cleaner. The capabilities of IoT devices can be represented in an attack graph in the same way that vulnerabilities are modeled in the traditional attack graph.

However, the IoT diversity make the representation of capabilities a challenging task, since two devices of the same type might have different capabilities. Moreover, a certain device and its capabilities may be exploited in different ways when deployed in different environment. For example, if an attacker can monitor the video stream of a security camera, the location and position of the camera affects the information she can obtain information. The solution need to support all the types of IoT devices and model each functionality for each device in each possible context.

## 5 Using Attack Graphs to Optimize Deployment of IoT Devices

In this section, we will present how an adjusted attack graph can solve the deployment challenge of IoT devices.

IoT devices can be deployed in a wide range of locations, but due to security risks and great diversity, their deployment need to be considered twice. Careless deployment of IoT devices can potentially help an attacker to penetrate a network. Adjusting attack graphs to support networks with IoT devices, might enable the graphs to supply the solution for the deployment challenge.

For example, in Figure 2 the camera's goal is to monitor the corridor, and could be deployed only at the left or right side of the corridor. If the camera is located on the left side of the corridor, it will connect to the Regular VLAN via L, and via R vice versa.

As presented in Figure 3, when the camera is deployed on the left side of the corridor there is only one path to the goal node. The attacker needs to infect the camera and vacuum cleaner in the morning and gain access to the Restricted network at night (when the vacuum cleaner changes its location). However, deploying the camera on the right side of the corridor is not secure, since it creates another attack path (as shown in Figure 4). In this path, which can be exploited at any time of day, the attacker can acquire P2 and combine it with E8 to connect the camera to the Restricted network (P5). Therefore, deploying the camera on the left side of the corridor is a more secure choice than deploying the camera on the right side.

Our goal is to find the optimal deployment of a set of IoT devices -- one that will minimize the security risk. Given a set of IoT devices and a list of constraints for each device (e.g., the camera must be deployed in the corridor), determining the optimal deployment can be accomplished by evaluating the risks to the network posed by every possible deployment. The main weaknesses of this approach are its time complexity and the need for reliable network scanning that represents the network, including the IoT devices, accurately.

# 6      Conclusion

In this paper, we demonstrated how IoT devices affect the security assessment of enterprise networks and create new challenges for state-of-the-art assessment methodologies based on attack graphs. We pointed out aspects that need to be addressed in each of the three main steps of the attack graph analysis methodology and demonstrated the challenges using a simple scenario.

We believe that in order to improve the security of enterprise organizations that deploy IoT devices on their premises, traditional security and assessment tools need to be adapted to address the challenges posed by the IoT devices' characteristics. In future work, we plan to implement our requirements using MulVal or another open source framework. In addition, we plan to investigate methods for risk assessment of IoT devices that consider a device's capabilities in specific contexts. Finally, we plan to address the IoT device deployment challenges by developing a method that can determine the most secure deployment of a set of IoT devices by using attack graphs.